\documentclass[twocolumn,amsmath,amssymb,aps]{revtex4}
\usepackage{booktabs}
\usepackage{graphicx}
\usepackage{xcolor}
\usepackage{longtable}
\usepackage{setspace}
\usepackage{array}

\begin{document}

\title{Partially deorbitalized meta-GGA}

\author{P. Bonf\`a}
\affiliation{Dipartimento di Scienze Matematiche, Fisiche, e Informatiche,
 Universit\`{a} di Parma, I-43124 Parma, Italy}
\author{S. Sharma}
\affiliation{Max-Born-Institute for Non-linear Optics and Short Pulse Spectroscopy,
 Max-Born Strasse 2A, 12489 Berlin, Germany}
\author{J. K. Dewhurst}
\affiliation{Max-Planck-Institut f\"{u}r Mikrostrukturphysik,
 Weinberg 2, D-06120 Halle, Germany}
\date{\today}

\begin{abstract}
Mejia-Rodriguez and Trickey recently proposed a procedure for removing
the explicit dependence of meta-GGA exchange-correlation
energy functionals $E_{\rm xc}$ on the kinetic energy density $\tau$.
We present a simple modification to this approach in which the
exact Kohn-Sham $\tau$ is used as input for
$E_{\rm xc}$ but the functional derivative of $\tau$ with respect to the
density $\rho$, required to calculate the potential term
$\int d^3r'\,\delta E_{\rm xc}/\delta\tau({\bf r}')|_{\rho}\cdot
 \delta\tau({\bf r}')/\delta\rho({\bf r})$,
is evaluated using an approximate kinetic energy density functional.
This ensures that the Kohn-Sham potential is a local multiplicative function as
opposed to the non-local potential of a generalized Kohn-Sham approach.
Electronic structure codes can be easily modified to use the new method.
We validate it by quantifying the accuracy of the predicted lattice
parameters, bulk moduli, magnetic moments and cohesive energies of a large
set of periodic solids. An unanticipated benefit of this method is to gauge
the quality of approximate kinetic energy functionals by checking if the
self-consistent solution is indeed at the variational minimum.
\end{abstract}

\maketitle
\section{Introduction}
The exchange-correlation energy functional, $E_{\rm xc}$, is an essential
component of the Kohn-Sham (KS)\cite{Kohn1965} formulation of
density functional theory (DFT)\cite{Hohenberg1964}.
Knowledge of $E_{\rm xc}$ allows for both the exact electronic density and the
total energy of a system of interacting electrons in an external potential
to be determined.
In practice however, $E_{\rm xc}$ has to be approximated, the inevitable result
of which is inexact densities and energies. Consequently, much effort has been
expended over that past several decades on inventing ever more sophisticated
approximations. 

The simplest approximation for the unknown exchange and correlation functional is the 
Local Density Approximation (LDA), in which the functional is supposed to depend locally on the charge density.
LDA represents the zeroth order expansion of $E_{\rm xc}$ in terms of electron density gradients and constitutes the first rung of the so-called `Jacob's ladder' of density functional
approximations\cite{Perdew2001}.
On the second rung, the Generalized Gradient Approximation (GGA), takes
into account density gradients\cite{Perdew1996}, and satisfies more known properties of the exact functional.
However, because of the limitations
of a semi-local functional form, GGAs tend to be accurate for either
energies or equilibrium geometries, but not both\cite{Perdew1998}.
The third rung of functionals,
meta-GGAs, were conceived to address this limitation
by including the Kohn-Sham kinetic energy (KE) density explicitly in their
formulation\cite{Becke1986,Becke1989,Tao2003}. This class of
functionals is truly
non-local because of the implicit dependence of the KE density
on the density itself.
This additional flexibility allows for more known exact conditions to be
satisfied. In fact, the strongly constrained and appropriately normalized (SCAN)
functional\cite{Sun2015,Sun2016} satisfies all 17 known exact conditions that
a meta-GGA can.

For the spin unpolarized case, the meta-GGA exchange-correlation
energy is given by
\begin{align*}
 E_{\rm xc}[\rho,\tau]=\int d^3r\,
 \epsilon_{\rm xc}\big(\rho({\bf r}),\tau({\bf r})\big)\rho({\bf r}),
\end{align*}
where
\begin{align}\label{rho_orb}
 \rho({\bf r})\equiv\sum_{i=1}^N
 \big|\varphi_i({\bf r})\big|^2
\end{align}
is the density and
\begin{align}\label{tau_orb}
 \tau({\bf r})\equiv\frac{1}{2}\sum_{i=1}^N
 \big|\nabla\varphi_i({\bf r})\big|^2
\end{align}
is the non-interacting, Kohn-Sham KE density,
with $\varphi_i$ being the $i$th Kohn-Sham orbital of a state
with $N$ electrons.
For its intended use, the KE density is not an independent variable
but rather an implicit functional of the
density, i.e. $\tau({\bf r})\equiv\tau[\rho]({\bf r})$.
Thus a difficulty arises when one has to determine the exchange-correlation potential $v_{\rm xc}$ as the functional
derivative of $E_{\rm xc}$ with respect to the density:
\begin{align}\label{vxc_mgga}
 v_{\rm xc}({\bf r})=
 \left.\frac{\delta E_{\rm xc}[\rho,\tau]}
 {\delta\rho({\bf r})}\right|_{\tau}
 +\int d^3r'\left.\frac{\delta E_{\rm xc}[\rho,\tau]}
 {\delta\tau({\bf r}')}\right|_{\rho}
 \frac{\delta\tau({\bf r}')}{\delta\rho({\bf r})}.
\end{align}
The last term requires the functional derivative of $\tau$ with respect to
$\rho$. This is numerically difficult to perform and requires an approach
similar to that used for the optimized effective potential (OEP)
method\cite{Yang16}. Instead, codes
typically calculate potentials determined from the derivative with respect to the
orbitals $\delta E_{\rm xc}/\delta\varphi({\bf r})$. Such an approach, however,
produces a non-local potential\cite{Zahariev2013} and is referred to as
{\em generalized} Kohn-Sham (gKS).

This presents us with a dilemma: the great effort expended to satisfy as
many exact constraints as possible is undermined by violating a fundamental
property of the KS potential, namely that it be a local function
in ${\bf r}$.
And yet, the choice of $\tau$ by the inventors of meta-GGA forces the writers
of electronic structure codes to have to deal with the difficult functional
derivative, or avoid it altogether with gKS.

Mej\'{i}a-Rodr\'{i}guez and Trickey (MRT) neatly sidestepped this problem
by replacing
the $\tau$ determined from the orbitals via Eq. (\ref{tau_orb}) with one
obtained from an approximate KE density
functional\cite{MejiaRodriguez17,MejiaRodriguez18,MejiaRodriguez2020}.
This `deorbitalized' meta-GGA was found to produce results of
accuracy which were comparable to that of the gKS method.
This approach however, removes the true non-locality of $E_{\rm xc}$ and
in effect reduces meta-GGA to a semi-local GGA-like functional
(albeit possibly with Laplacian terms\cite{Perdew2007}).

In the current work we adopt a `half-way' strategy, in that we use the
KS orbital-derived $\tau$ as {\em input} to the functional, but use an
approximate KE density functional to
evaluate the functional derivative in Eq. (\ref{vxc_mgga}). We term this
approach `partial deorbitalization' and find that even a fairly primitive
KE functional, like the Thomas-Fermi-von Weizs\"{a}cker (TFvW)
gradient expansion\cite{Yang86}, yields accurate results. This deorbitalization
scheme can be easily implemented in codes which already employ the
generalized Kohn-Sham version of meta-GGA.
Partial deorbitalization retains the `exact' $\tau$ for the energy but
utilizes both the exact and approximate $\tau$ for the potential, which favors situations
in which the main error is functional-driven rather than
density-driven\cite{Medvedev2017}.

\section{Approximations}
As will be demonstrated later, the method does not require a particularly
sophisticated KE density functional for calculating the
functional derivative
$\delta\tau({\bf r}')/\delta\rho({\bf r})$ in Eq. (\ref{vxc_mgga}).
Here we choose to use the gradient expansion of $\tau$ with the
TFvW terms\cite{Yang86} for the sake of ease of implementation:
\begin{align}
 \tilde{\tau}(\rho({\bf r}),\sigma({\bf r}))=
 \frac{3}{10}(3\pi^2)^{2/3}\rho^{5/3}({\bf r})
 +\frac{1}{72}\frac{\sigma({\bf r})}{\rho({\bf r})},
\end{align}
where $\sigma({\bf r})\equiv|\nabla\rho({\bf r})|^2$.
In this case, the second term in Eq. (\ref{vxc_mgga}) becomes
\begin{align}
 \int d^3r'\,w_{\rm xc}({\bf r}')&
 \frac{\delta\tilde{\tau}({\bf r}')}{\delta\rho({\bf r})}
 =w_{\rm xc}({\bf r})\left.\frac{\partial\tilde{\tau}({\bf r})}
 {\partial\rho({\bf r})}\right |_{\sigma}\nonumber \\
 &-2\nabla\cdot\left[w_{\rm xc}({\bf r})
 \left.\frac{\partial\tilde{\tau}({\bf r})}
 {\partial\sigma({\bf r})}\right|_{\rho}\nabla\rho({\bf r})\right],
\end{align}
where
$w_{\rm xc}({\bf r})\equiv\delta E_{\rm xc}/\delta\tau({\bf r})|_{\rho}$.

The spin polarized case is a straight-forward extension to the
unpolarized case: the meta-GGA functional is generalized to the collinear form
\begin{align*}
 E_{\rm xc}[\rho^{\uparrow},\rho^{\downarrow},\tau^{\uparrow},\tau^{\downarrow}]
 =\int d^3r\,
 \epsilon_{\rm xc}\big(\rho^{\uparrow}({\bf r}),\rho^{\downarrow}({\bf r}),
 \tau^{\uparrow}({\bf r}),\tau^{\downarrow}({\bf r})\big)\rho({\bf r}),
\end{align*}
where $\rho^{\uparrow\downarrow}$ and $\tau^{\uparrow\downarrow}$ are the
up- and down-spin density and KE density, respectively.
The total Kohn-Sham KE satisfies\cite{Oliver1979}
\begin{align*}
 T_s[\rho^{\uparrow},\rho^{\downarrow}]
 =\tfrac{1}{2}T_s[2\rho^{\uparrow}]+\tfrac{1}{2}T_s[2\rho^{\downarrow}],
\end{align*}
thus we will take spin-up KE density to depend
exclusively on the spin-up density\cite{Perdew2007},
$\tau^{\uparrow}[\rho^{\uparrow},\rho^{\downarrow}]({\bf r})\equiv
 \tau^{\uparrow}[\rho^{\uparrow}]({\bf r})$,
and likewise for the spin-down density.
This implies that
$\delta\tau^{\uparrow}({\bf r}')/\delta\rho^{\downarrow}({\bf r})
 |_{\rho^{\uparrow}}=
 \delta\tau^{\downarrow}({\bf r}')/\delta\rho^{\uparrow}({\bf r})
 |_{\rho^{\downarrow}}=0$.
The spin-up exchange-correlation potential, for example, is then given by
\begin{align*}
 v_{\rm xc}^{\uparrow}({\bf r})=
 \left.\frac{\delta E_{\rm xc}}{\delta\rho^{\uparrow}({\bf r})}
 \right|_{\rho^{\downarrow},\tau^{\uparrow},\tau^{\downarrow}}
 +\int d^3r'\left.\frac{\delta E_{\rm xc}}{\delta\tau^{\uparrow}({\bf r}')}
 \right|_{\rho^{\uparrow},\rho^{\downarrow},\tau^{\downarrow}}
 \frac{\delta\tau^{\uparrow}({\bf r}')}
 {\delta\rho^{\uparrow}({\bf r})},
\end{align*}
which can be easily evaluated for the spin-polarized TFvW KE density
\begin{align*}
 \tilde{\tau}^{\uparrow}(\rho^{\uparrow}({\bf r}),
 \sigma^{\uparrow}({\bf r}))=
 \tfrac{1}{2}\tilde{\tau}(2\rho^{\uparrow}({\bf r}),
 4\sigma^{\uparrow}({\bf r})),
\end{align*}
where $\sigma^{\uparrow}({\bf r})=|\nabla\rho^{\uparrow}({\bf r})|^2$.

\section{Computational Details}
The goal of the following section is to evaluate the accuracy of our partial
deorbitalization strategy. We do so by validating its {\it ab initio}
predictions against a large number of experimental results. We also list the
computational outcomes of the previous work by
MRT\cite{MejiaRodriguez18}, including both the results from their gKS
and the fully deorbitalized meta-GGA simulations, in order to
compare against different strategies for deorbitalizing meta-GGA.
Notably, the comparison among these approximations, i.e. our work and the
findings of MRT, must be performed `cum grano salis', since the simulations
have been carried out using slightly different conditions.
MRT\cite{MejiaRodriguez18} used the plane-wave based code
VASP\cite{Kresse1994,Kresse1996},
adopting PAW pseudopotentials\cite{Kresse1999},
and the SCAN exchange and correlation functional\footnote{A complete description
of the methodological details is provided in the original reference.}. In what
follows we will call this combination gKS-SCAN when referring to
simulations performed with the gKS, while the label FD-SCAN
will indicate their fully deorbitalized results.
In our work we opt instead for the rSCAN functional\cite{Bartok2019} in order to
overcome convergence problems with full potential (FP) simulations. 
The partially deorbitalized method introduced above will therefore be labelled
PD-rSCAN. The label should immediately remind the reader of the two main
ingredients to be considered: the deorbitalization scheme (if any) and the
choice of the functional providing the exchange and correlation contribution.
Finally, we also performed a number of simulations within the gKS scheme
and we refer to these results as gKS-rSCAN, 
to distinguish them from the gKS results of MRT, labeled gKS-SCAN.

In order to compute equilibrium lattice parameters and bulk moduli with high
accuracy, we opt for a FP description of the electronic wave-functions with an
APW basis\cite{Singh2006}. This choice is effective for periodic systems
but makes it difficult
to compute the total energy of isolated atoms. To overcome this problem, we
also use a plane-wave basis set and pseudopotentials for the estimation of the
cohesive energies.

The Elk code\cite{elk} version 8.3.15 is used to perform FP simulations.
Reciprocal space sampling is performed with at least a $17 \times 17 \times 17$
Monkhorst-Pack grid.
The basis is expanded up to $2|{\bf G}+{\bf k}|_{\rm max}\geq 8$. The
remaining parameters are set by the {\tt vhighq} option.

For plane-wave simulations we use the Quantum ESPRESSO package \cite{QE-2009}
and opt for
norm-conserving pseudopotentials \cite{Schlipf2015} generated with the PBE \cite{Perdew1996} exchange
and correlation functional. The reciprocal space sampling and the cut-off
energy for the KS wavefunction expansion have been converged in order to
obtain better than 1 mRy/atom accuracy in the total energy.


Estimation of atomic energies with meta-GGA requires further care.
The exponentially vanishing charge produces problematic behavior in the exchange
and correlation potential across the self consistent cycles. There are two
options to improve the convergence.
The first is to converge isolated atom simulations with GGA (we used PBE) and later reuse the converged
electronic charge as the starting point of meta-GGA simulations
(for both gKS-rSCAN and PD-rSCAN).
The alternative method consists of introducing a cut-off for vanishing
charge density that removes the ill-behaving contributions from the exchange
and correlation potential. This parameter can be converged together with the
remaining settings governing the basis expansion. Numerically equivalent
results are obtained with both approaches, when convergence can be achieved.
The details are reported in the Supplemental Information.

Finally, we point out that a more recent analysis\cite{MejiaRodriguez2020} of
deorbitalized meta-GGA adopts r$^2$SCAN\cite{Furness2020} and the authors
report that the convergence issues discussed in their previous
work\cite{MejiaRodriguez18} appear to be due to the functional itself rather
than to their deorbitalization strategy. We tried r$^2$SCAN but still found
difficulties converging FP simulations and therefore, in order to preserve
consistency among our PW and FP results, we abandoned this option.

Following MRT, the equilibrium lattice constants $a_0$ and bulk moduli $B_0$ at
$T = 0$ K were determined by calculating the total energy per unit cell in the
range $V_0 \pm 10$\% (where V0 is the equilibrium unit cell volume), followed
by a twenty point fit to the stabilized jellium equation of state
(SJEOS)\cite{MejiaRodriguez18, Alchagirov2001}.

\begin{figure}
  \centering
  \includegraphics{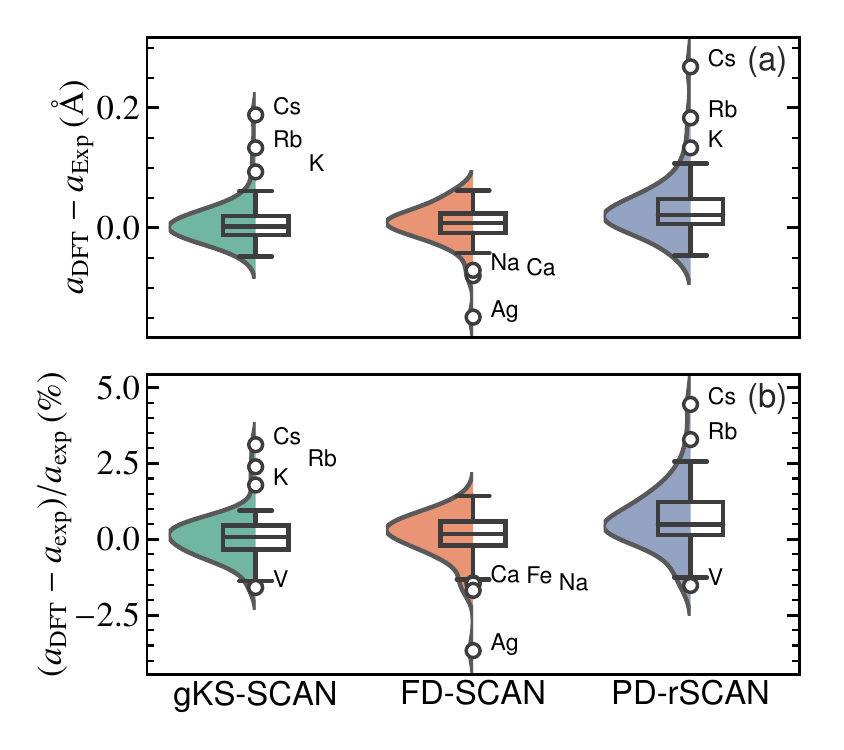}
  \caption{Visualization (violin plot and box plot) of the deviations from
   experimental data of the predicted lattice parameter obtained with the
   different implementations of meta-GGA. The upper(lower) panel gives
   absolute(relative) deviations. The violin plots (transparent color) represent
   the data distribution and are based on a Gaussian kernel density estimation
   implemented in seaborn\cite{Waskom2021}.
   In the box plot, the boxes hold 50\% of the data,
   with equal number of data points above and below the median deviation
   (full black line). The whiskers indicate the range of data falling within
   1.5$\times$box-length beyond the upper and lower limits of the box (from the
   first quartile to the third quartile). The whiskers extend from the box by
   1.5$\times$ the inter-quartile range. Outliers beyond this range are indicated with
   circular makers and the solids' labels are reported on the right of each
   point.\label{fig:latt}}
\end{figure}

\begin{figure}
  \centering
  \includegraphics{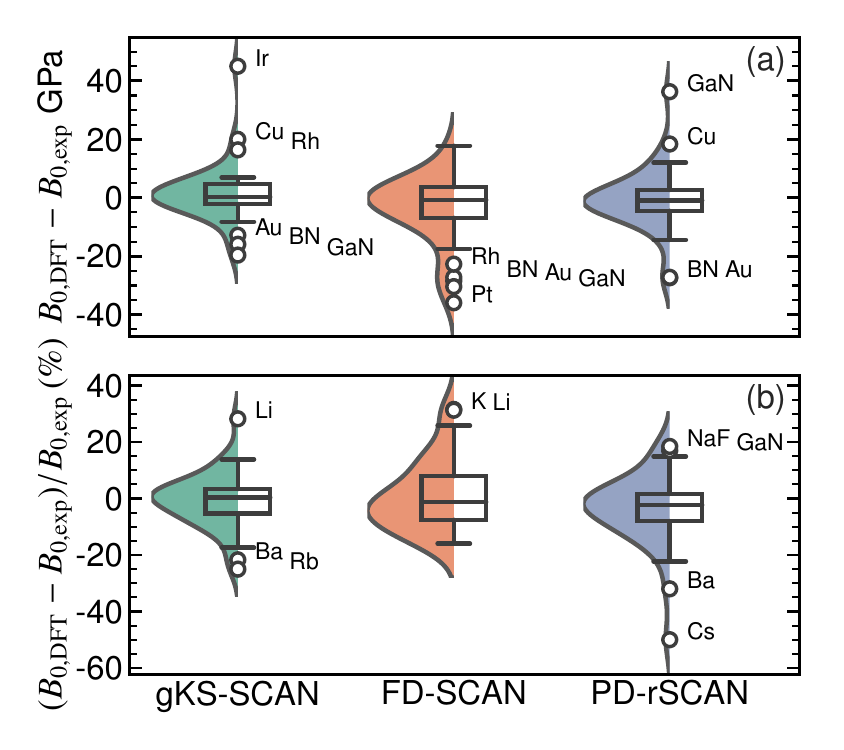}
  \caption{Visualization (violin plot and box plot) of the deviations from
   experimental data of the predicted bulk modulus obtained with the different
   implementations of meta-GGA. See Fig. \ref{fig:latt} and main text for details.}
  \label{fig:b0}
\end{figure}

\section{Results and Discussion}
The complete set of results obtained for the computation of the equilibrium
lattice parameters of 51 solids is presented, together with statistical data analysis, in Fig.~\ref{fig:latt}, where panels (a) and (b) report the absolute and relative deviations.
All absolute values are reported in Table SIII of the supplemental material (SM) \footnote{See Supplemental Material at [URL will be inserted by publisher] for the full list of computational results and the details of plane wave based simulations.}. Values
for gKS-SCAN and FD-SCAN are taken from Ref.~\cite{MejiaRodriguez18} while our
results for PD-rSCAN are shown in the third column of the figure and in the fifth column of the table.
Finally, the reference experimental data are based on
zero-point corrected experimental lattice constants, detailed in
Ref. \cite{Tran2016} and references therein.


FD-SCAN and PD-rSCAN show similar trends and comparable agreement with respect
to experimental lattice parameters. Alkali metals are a notable exception, where
deviations by more than 2\% \footnote{The equilibrium lattice parameter for Cs
evaluated with the gKS with QE results in 6.26 \AA, at odd with the prediction
obtained with VASP by MRT, i.e. 6.09 \AA} are observed. This is indeed a known
issue with SCAN that originates from a poor description of the semi-core region
of these elements, as explained in detail in Ref. \cite{Kovacs2019}. 
No clear indications of over-binding or
under-binding can be identified when considering at the whole set. Yet we note
that FD-SCAN and PD-rSCAN show similar trends across the periodic table.
The mean absolute deviation (MAD) is 0.028 \AA, 0.026 \AA{} and 0.036 \AA{} for
gKS-SCAN, FD-SCAN and PD-rSCAN respectively.
While the distribution of deviations for PD-rSCAN is slightly broader than the
other two, it is noted that outliers are found in the upper part of the box-plot
for both gKS-SCAN and PD-rSCAN meta-GGA, while the fully deorbitalized approach
of MRT behaves differently, showing only outliers on the opposite side of the other two distributions.

Fig. \ref{fig:b0} reports the same statistical analysis for the bulk moduli of various cubic and hexagonal systems. All results are also tabulated in Table SIV of the SM.
The most significant discrepancies are observed also in this case for
alkali metals. Additional outliers are found in transition metal elements, Cu
and Au, and in wide-bandgap semiconductors GaN and BN.
The distribution of PD-rSCAN and gKS-SCAN compare equally well
against the experiment, while FD-SCAN is showing slightly worse performance, as it can be appreciated from the box-plots of Fig.~\ref{fig:b0}.
The MAD for the bulk modulus is 6.8, 9.4 and 6.7 GPa for gKS-SCAN,  FD-SCAN and
PD-rSCAN respectively. As expected, our approach gives bulk moduli that are
similar to those gKS-SCAN, but the partially deorbitalization formulation
shows slightly improved values on average with respect to FD-SCAN.

\begin{table} 
 \caption{Calculated Kohn-Sham band gaps in eV for 20 insulators or
 semiconductors in the test set. The last column reports optimized effective
 potential results from Yang {\it et al.} Ref. \cite{Yang2016}
 obtained with the Krieger-Li-Iafrate approximation\cite{Krieger1992}.\label{tab:gaps}}
 \begin{tabular}{cccccc}
 \hline\hline
 Solid       & Expt.  & SCAN   & FD-SCAN & PD-rSCAN & KLI (Ref. \cite{Yang2016}) \\
 \hline
  C          &  5.50  & 4.54   &  4.22  & 3.92 & 4.26 \\ 
  Si         &  1.17  &  0.82  &  0.80  & 0.68 & 0.78 \\ 
  Ge         &  0.74  &  0.14  &  0.00  & 0.025 & --- \\ 
  SiC        &  2.42  &  1.72  &  1.55  & 1.48  & --- \\ 
  BN         &  6.36  &  4.98  &  4.66  &  4.50 & 4.73 \\ 
  BP         &  2.10  &  1.54  &  1.41  &  1.32 & 1.52 \\ 
  AlN        &  4.90  &  3.97  &  3.50  &  3.47 & --- \\ 
  AlP        &  2.50  &  1.92  &  1.81  &  1.68 & --- \\ 
  AlAs       &  2.23  &  1.74  &  1.59  &  1.55 & --- \\ 
  GaN        &  3.28  &  1.96  &  1.49  &  1.53 & --- \\ 
  GaP        &  2.35  &  1.83  &  1.72  &  1.63 & 1.72 \\ 
  GaAs       &  1.52  &  0.77  &  0.33  &  0.48 & 0.45 \\ 
  InP        &  1.42  &  1.02  &  0.59  &  0.53 & 0.77 \\ 
  InAs       &  0.42  &  0.00  &  0.00  &  0.30 & --- \\ 
  InSb       &  0.24  &  0.00  &  0.00  &  0.04 & --- \\ 
  LiH        &  4.94  &  3.66  &  3.69  &  3.48 & --- \\ 
  LiF        &  14.20 &  10.10 &  9.16  &  9.36 & 9.11 \\ 
  LiCl       &  9.40  &  7.33  &  6.80  &  6.59 & --- \\ 
  NaF        &  11.50 &  7.14  &  6.45  &  6.43 & --- \\ 
  NaCl       &  8.50  &  5.99  &  5.59  &  5.31 & 5.25 \\ \hline\hline
\end{tabular}
\end{table}

The Kohn-Sham (KS) band gaps for selected insulators and semiconductors are
shown in Table \ref{tab:gaps}. The results obtained with PD-rSCAN are similar
to the ones produced by FD-SCAN, and in both cases the values are smaller
than those obtained with gKS-SCAN. This systematic difference is a well known property of the generalized KS
theory and an accurate analysis of this point is presented in Ref. \cite{Yang2016}.
The results of FD-SCAN and PD-rSCAN are instead obtained with a local potential
and are therefore expected to match the previous results of the optimized
effective potential reported in the last column of Table \ref{tab:gaps}.

\begin{table} 
 \caption{Plane wave results: calculated cohesive energies in eV/atom.
 PBE pseudopotentials are used in this case. FD-SCAN data are from
 Ref. \cite{MejiaRodriguez18} and have been obtained with a different set
 of pseudopotentials.\label{tab:cohesion3}. Experimental values are from Ref.~\cite{PhysRevX.6.041005,kittel1996introduction} and references therein.}
 \begin{tabular}{cccccc}
 \hline\hline
 & Expt. & \multicolumn{2}{c}{MRT} & \multicolumn{2}{c}{This work} \\
 \hline
 &     &  gKS-SCAN &  FD-SCAN &  gKS-rSCAN &  PD-rSCAN \\
 \hline
 Al &   3.43 &  3.57 &  3.52 &  3.78 &  3.79 \\
 Ag &   2.96 &  2.76 &  2.65 &  2.86 &  2.87 \\
 Ba &   1.91 &  1.48 &  1.96 &  1.91 &  1.90 \\
 Be &   3.32  &  ---  &  ---  &  4.04 &  4.06 \\
 Ca &   1.87 &  1.87 &  1.98 &  2.04 &  2.03 \\
 Ge &   3.89 &  3.94 &  3.82 &  4.16 &  4.18 \\
 Ir &   6.99 &  7.08 &  6.80 &  7.98 &  8.00 \\
  K &   0.94 &  0.81 &  0.78 &  0.92 &  0.84 \\
 Na &   1.12 &  1.04 &  0.99 &  1.12 &  1.13 \\
 Pd &   3.93 &  4.16 &  4.07 &  4.16 &  4.16 \\
 Rh &   5.78 &  5.22 &  5.65 &  6.29 &  6.32 \\
 Ru &   6.77 &  6.23 &  6.31 &  7.38 &  7.40 \\
 Si &   4.68 &  4.69 &  4.60 &  4.74 &  4.75 \\
 \hline\hline
 \end{tabular}
\end{table}

Cohesive energies are reported in Table \ref{tab:cohesion3} along
with experimental atomization energies from Ref. \cite{MejiaRodriguez18, kittel1996introduction, PhysRevX.6.041005}.
As already mentioned, these results have been obtained with PW based simulations
and in this case a larger discrepancy between the {\it ab initio} predictions
and the experiment is observed, especially for transition metals. Notably, MRT
results obtained with the gKS-SCAN approach are in slightly better agreement than our gKS-rSCAN (first and third columns). This
may be due to the norm conserving pseudopotentials used in our simulations that
have been generated with a different functional (PBE) for the core electrons
and miss the KE density contribution.
The comparison between PD-rSCAN and gKS simulations (third and fourth columns)
is instead showing perfect agreement, with the only exception being K.

\begin{table}
 \caption{Calculated magnetic moment (in $\mu_B$) for a selection of
 ferromagnets and anti-ferromagnets. Experimental results are
 taken from Ref. \cite{Tran2020} and references therein.\label{tab:moms}}
 \begin{tabular}{ccccc}
 \hline\hline
 Solid & Expt. & SCAN & FD-SCAN & PD-rSCAN \\
 \hline
 \multicolumn{5}{c}{Ferromagnets} \\
 \hline
  Fe   & 1.98-2.08 &  2.60  &  2.05  &  2.23 \\
  Co   & 1.52-1.62 &  1.80  &  1.63  &  1.69 \\ 
  Ni   & 0.52-0.55 &  0.78  &  0.67  &  0.67 \\ 
  V    & 0.00      &  0.57  &  0.0   & 0.0 \\
 \hline
 \multicolumn{5}{c}{Anti-ferromagnets}   \\ \hline
  FeO  & 3.32-4.6  & 3.60  &  ---   &  3.51 \\ 
  CoO  & 3.35-3.98 & 2.61  &  ---   &  2.5 \\ 
  NiO  & 1.9-2.2   & 1.61  &  ---   &  1.44 \\
  MnO  & 4.58      & 4.52  &  ---   &  4.40 \\
 \hline\hline
 \end{tabular}
\end{table}

\begin{figure}[ht]
\includegraphics[width=\columnwidth]{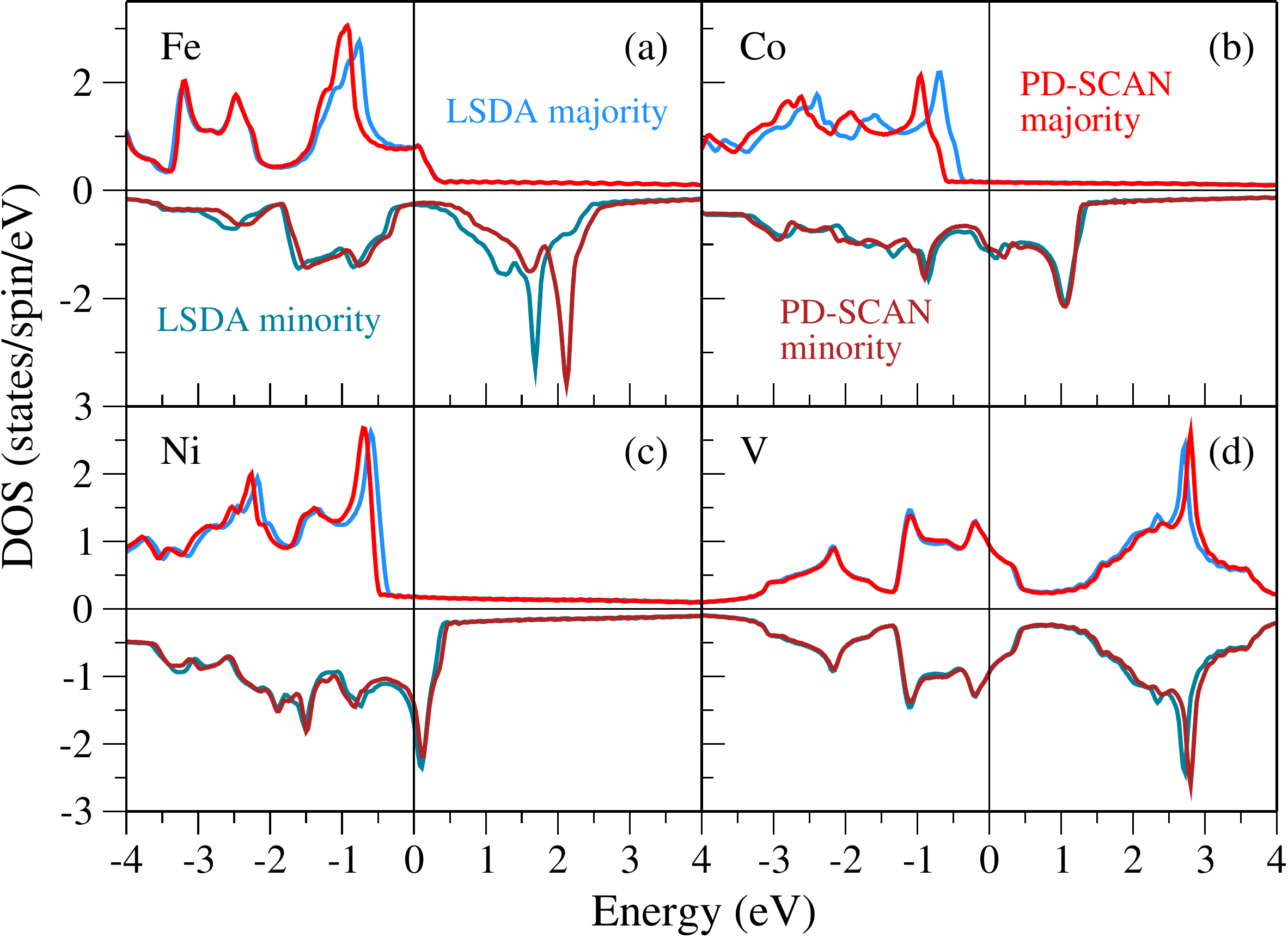}
\caption{Spin projected density of states (DOS) in states/spin/eV for
 (a) Fe (b) Co (c) Ni and (d) V. Results are shown
 for LDA as well as PD-rSCAN.}\label{fig:dos}
\end{figure}

Magnetic properties, reported in Table \ref{tab:moms}, are the most sensitive to
the choice of the deorbitalization scheme. It has indeed already extensively
discussed how gKS with SCAN leads to overhestimated magnetization in
transition metal elements\cite{MejiaRodriguez2019, Tran2020, Fu2019}.
On the other hand, FD-SCAN and PD-rSCAN improve the agreement with
experimental results for the elemental
ferromagnets Fe, Co and Ni, and also predict the expected
non-magnetic ground state for vanadium.
The density of states for the four elemental solids obtained with LDA
and PD-rSCAN is shown in Fig. \ref{fig:dos}.
Small differences in the densities of states of the ground states can be
appreciated. The plot shows that, relative to LDA, PD-rSCAN shifts
the spin majority occupied states downward, while the spin minority state
energies are only slightly increased in all elements but iron, where the effect
is more pronounced but only in the conduction bands.
The overall effect is very limited and indeed the resulting magnetic moments
are very close.

For the anti-ferromagnetic, insulating magnetic oxides FeO, CoO, NiO and MnO
the picture is more mixed. The atomic moment of MnO is close to the experimental
value and that of FeO lies within the admittedly broad range of measured moments.
However, the moments of CoO and NiO are underestimated by both FD-SCAN and PD-rSCAN.
This has been attributed to strong correlation effects which are not fully
described even by meta-GGA functionals.

\begin{figure}[ht]
\includegraphics[width=\columnwidth]{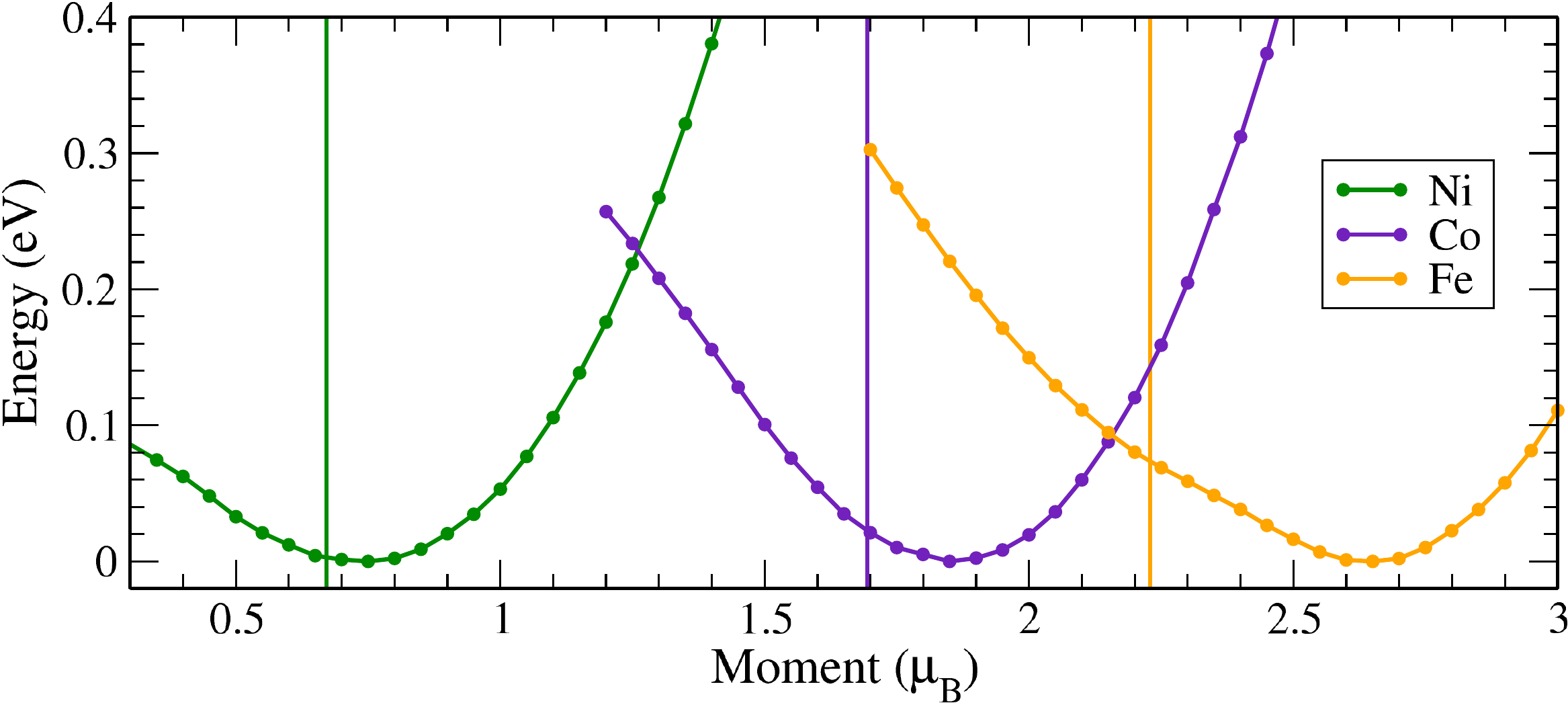}
\caption{Energy vs. moment graphs for Fe, Co and Ni using the FP code with
 PD-rSCAN. The vertical lines
 indicate the moments obtained self-consistently.}\label{fig:e_vs_m}
\end{figure}

By calculating the total energy while keeping the moment constrained
to a given value (referred to as a fixed spin-moment calculation), it can
be ascertained if the self-consistent solution is truly variational.
The energy vs. moment for Fe, Co and Ni is plotted in Fig. \ref{fig:e_vs_m}
along with the moment from the corresponding
self-consistent solutions. As can be seen,
the moments are generally smaller than the location of the minima of the curves.
This implies that the calculations are not perfectly variational, which in
turn implies, unsurprisingly, that the approximate KE functional is inexact.
Fully deorbitalized functionals will not suffer from this inconsistency because
the functional derivative is determined from same KE functional as is used to
evaluate the energy (inexact though it is). Nevertheless, this mismatch does
serendipitously present us with
a useful tool for determining the accuracy of approximate KE functionals:
simply check if the solution is at the energy minimum. This is particularly
useful because these calculations are for real-world atomistic systems as
opposed to simplified models. Magnetic moments can be used as done here,
but there are other possibilities: for example testing if the Helmann-Feynman
forces on the nuclei are strictly zero at the lowest energy.

\section{Conclusions}
We have modified the deorbitalized approach introduced by MRT by retaining
the true Kohn-Sham KE density as input to any meta-GGA
exchange-correlation functional,
while employing an approximate KE functional for the functional derivative
of $\tau$ with respect to $\rho$. This represents a
simple route to keeping both $E_{\rm xc}$ inherently non-local and
$v_{\rm xc}$ as a local operator in ${\bf r}$.
DFT codes which already have the gKS version of meta-GGA implemented
can be easily modified to accommodate this scheme.
We find that even a relatively crude KE functional like TFvW yields results
which are, on the whole, at least as accurate as competing approaches such
as generalized Kohn-Sham or full deorbitaliztion.
Furthermore, there is ample scope for improvement with respect to both
$E_{\rm xc}$ as well as the KE functional. For example,
KE functionals involving the Laplacian of the
density\cite{Oliver1979,Perdew2007} could
further enhance the accuracy of method. Showing that the self-consistent
solution does
indeed correspond to the energy minimum for solids and molecules is a useful
measure of the quality of the KE functionals in real-world situations.
Lastly, our method may be extended to the case of non-collinear
exchange-correlation meta-GGA functionals, at least two of which have been
developed recently\cite{Pu2023,TancogneDejean2022}.
Treating this type of functional with partial deorbitalization will
also require a generalization of the spin-dependent KE density functional to
the non-collinear case\cite{Tellgren2018}.

\section{Acknowledgements}
PB thanks Andrea Ferretti and Pietro Delugas for fruitful discussions and the students of a work-related internship (Italian PCTO activity) Giacomo Bonvicini, Filippo Pedrazzani, Niccolò Lo Re, Alessandro Zanichelli and Giulio Cacciapuoti for analyzing the equation of state of various elements reported in the manuscript.

\bibliography{bibliography}


\clearpage

\setcounter{equation}{0}
\setcounter{figure}{0}
\setcounter{table}{0}
\renewcommand{\theequation}{S\arabic{equation}}
\renewcommand{\thefigure}{S\arabic{figure}}
\renewcommand{\thetable}{S\arabic{table}}
\renewcommand{\bibnumfmt}[1]{[S#1]}

\section*{Supplemental Material}
\subsection*{Convergence of plane wave based meta-GGA simulations}
\begin{figure}
    \centering
    \includegraphics{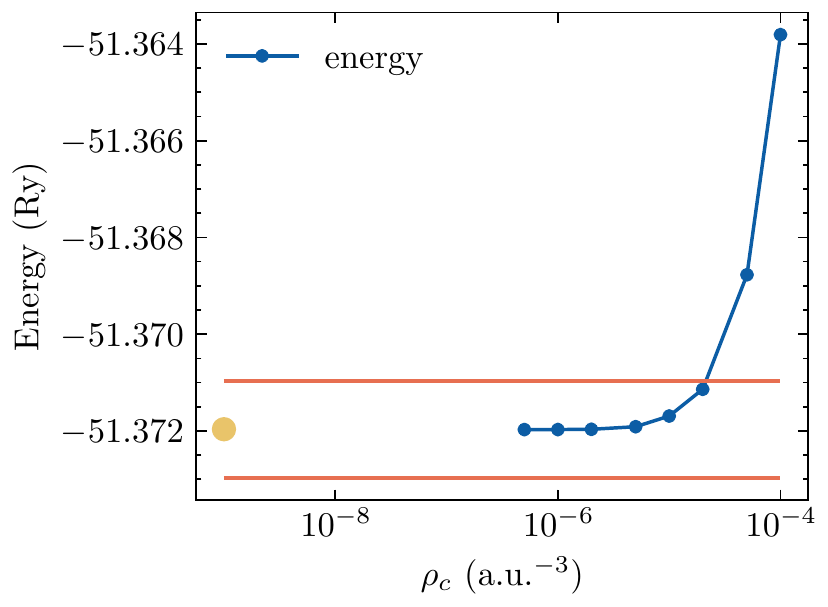}
    \caption{Convergence of total energy of isolated Ba atom against charge threshold $\rho_c$. The red lines highlight 1~mRy accuracy. The yellow point shows the result obtained by using converged charge density obtained with PBE as the starting point for the meta-GGA simulation.}
    \label{fig:barhocut}
\end{figure}

The estimation of atomic energies with meta-GGA is not straightforward and requires the adoption of workarounds to the pathological behaviour of the rSCAN functional with vanishing charges.
We found two options:
1. converge the isolated atom states with PBE and use the resulting density as an input for the meta-GGA simulation, or
2. introduce a cutoff such that
\begin{equation}
 v_{xc}^{\text{cut}}(\mathbf r) = \begin{cases}
     v_{xc}(\mathbf r) & \text{if } \rho(\mathbf r) > \rho_c \\
     0 & \text{otherwise}
\end{cases}   
\end{equation}
We then converge the total energy against the box size, the plane wave cutoff $E_c$, the charge density expansion (which is larger than $4 E_c$ in order to achieve numerical stability of meta-GGA) and, when using $ v_{xc}^{\text{cut}}$, against $\rho_c$.
An example of convergence is shown in Fig.~\ref{fig:barhocut}. When both method 1. and 2. achieve convergence, the results are found to be numerically equivalent.

The cutoff for charge expansion is set to 18 times $E_c$ and a total energy convergence better than 1~mRy is achieved with the following parameters for the atoms listed in the main text ( box size in \AA{}, $E_c$ in Ry): (80, 10) for Al, Ag, Ru; (70, 10) for Ba, Ca; (70, 12) for Pd, Ge; (110, 10) for Be; (70, 14) for K, Na; (50, 10) for Si, Ir; (80, 14) for Rh.

The estimation of the equilibrium volume of the above elements arranged in periodic solids is not problematic but we stress that unexpectedly large thresholds for the charge density expansion are required to converge the total energy of rSCAN simulations and obtain smooth $E$ vs $V$ curves.
The values used for the elemental solids listed above is reported in Tab.~\ref{tab:pwconv}. Monkhorst-Pack grids were used to sample reciprocal space and a Methfessel-Paxton smearing of 0.01~Ry was adopted for metals.

\LTcapwidth=\columnwidth
\renewcommand*{\arraystretch}{1.5}
\begin{longtable}{cccc}
\hline
        Compound & $E_c$ & $E_\rho$ & $k$-grid \\
        \hline
         Al & 80 & 1440 & (14, 14, 14) \\
         Ag & 90 & 1080 & (16, 16, 16) \\
         Ru & 50 & 600 & (15, 15, 15) \\
         Be & 70 & 840 & (9, 9, 9) \\
         Ca & 50 & 600 & (10, 10, 10) \\
         Pd & 90 & 1620 & (12, 12, 12) \\
         Ge & 90 & 1080 & (9, 9, 9) \\
         Ba & 50 & 600 & (9, 9, 9) \\
         K  & 60 & 720 & (7, 7, 7) \\
         Na & 100 & 2000 & (16, 16, 16) \\
         Si & 40 & 720 & (9, 9, 9) \\
         Ir & 70 & 1260 & (20, 20, 20) \\
         Rh & 50 & 900 & (9, 9, 9) \\
    \caption{Parameters used to compute $E$ vs $V$ curves with the plane wave basis. Energies are in Ry.\label{tab:pwconv}}

\end{longtable}

\subsection*{Tabulated results}
The results obtained for all materials of our testing set are reported in Tab.~\ref{tab:latpar1} and  Tab.~\ref{tab:bmod} and compared with the outcomes of the previous investigation by Mejia-Rodriguez and Trickey \cite{MejiaRodriguez18}.

\setlength{\extrarowheight}{0.01mm}
\begin{longtable}{ccccc}

 \caption{Experimental and calculated equilibrium lattice parameters (\AA{}) of
 51 solids. 
 \label{tab:latpar1}}\\
 \hline\hline
 Solid & Expt. & gKS-SCAN & FD-SCAN & PD-rSCAN \\
 \hline
 \endfirsthead
 \hline\hline
 Solid & Expt. & gKS-SCAN & FD-SCAN & PD-rSCAN \\
 \hline
 \endhead
 \hline
 \endfoot
 \hline\hline
 \endlastfoot
     Ag & 4.062 & 4.081 & 3.913 & 4.098 \\
     Al & 4.018 & 4.006 & 3.997 & 3.999 \\
   AlAs & 5.649 & 5.671 & 5.659 & 5.670 \\
    AlN & 4.368 & 4.360 & 4.364 & 4.371 \\
    AlP & 5.451 & 5.466 & 5.449 & 5.473 \\
     Au & 4.062 & 4.086 & 4.120 & 4.123 \\
     BN & 3.592 & 3.606 & 3.612 & 3.615 \\
     BP & 4.525 & 4.525 & 4.530 & 4.532 \\
     Ba & 5.002 & 5.034 & 5.027 & 5.109 \\
     Ca & 5.556 & 5.546 & 5.476 & 5.575 \\
     Co & 3.524 & 3.505 & 3.503 & 3.500 \\
     Cs & 6.039 & 6.227 & 6.090 & 6.307 \\
     Cu & 3.595 & 3.566 & 3.570 & 3.571 \\
     Fe & 2.853 & 2.855 & 2.811 & 2.817 \\
   GaAs & 5.640 & 5.659 & 5.677 & 5.667 \\
    GaN & 4.520 & 4.505 & 4.513 & 4.497 \\
    GaP & 5.439 & 5.446 & 5.445 & 5.443 \\
     Ge & 5.644 & 5.668 & 5.667 & 5.671 \\
     Hf & 3.151 & 3.123 & 3.159 & 3.197 \\
   InAs & 6.047 & 6.094 & 6.109 & 6.101 \\
    InP & 5.858 & 5.892 & 5.896 & 5.920 \\
   InSb & 6.468 & 6.529 & 6.528 & 6.560 \\
     Ir & 3.831 & 3.814 & 3.856 & 3.845 \\
      K & 5.212 & 5.305 & 5.238 & 5.345 \\
     Li & 3.443 & 3.457 & 3.470 & 3.510 \\
   LiCl & 5.070 & 5.099 & 5.086 & 5.091 \\
    LiF & 3.972 & 3.978 & 3.979 & 3.985 \\
    LiH & 3.979 & 3.997 & 3.969 & 3.990 \\
     Mo & 3.141 & 3.145 & 3.151 & 3.148 \\
     Na & 4.214 & 4.193 & 4.143 & 4.264 \\
   NaCl & 5.569 & 5.563 & 5.542 & 5.570 \\
    NaF & 4.582 & 4.553 & 4.574 & 4.601 \\
     Nb & 3.294 & 3.296 & 3.306 & 3.297 \\
     Ni & 3.508 & 3.460 & 3.488 & 3.465 \\
     Os & 2.699 & 2.686 & 2.710 & 2.738 \\
     Pd & 3.876 & 3.896 & 3.913 & 3.910 \\
     Pt & 3.913 & 3.913 & 3.956 & 3.941 \\
     Rb & 5.577 & 5.710 & 5.626 & 5.760 \\
     Re & 2.744 & 2.730 & 2.761 & 2.758 \\
     Rh & 3.794 & 3.786 & 3.817 & 3.799 \\
     Ru & 2.669 & 2.663 & 2.681 & 2.703 \\
     Sc & 3.270 & 3.271 & 3.261 & 3.324 \\
     Si & 5.421 & 5.429 & 5.423 & 5.429 \\
     Sr & 6.040 & 6.084 & 6.040 & 6.103 \\
     Ta & 3.299 & 3.272 & 3.300 & 3.302 \\
     Tc & 2.716 & 2.711 & 2.724 & 2.735 \\
     Ti & 2.915 & 2.897 & 2.898 & 2.929 \\
      V & 3.021 & 2.973 & 2.981 & 2.975 \\
      W & 3.160 & 3.149 & 3.165 & 3.169 \\
      Y & 3.594 & 3.608 & 3.599 & 3.674 \\
     Zr & 3.198 & 3.212 & 3.211 & 3.239 \\

\end{longtable}

\begin{longtable}{ccccc}
 \caption{Experimental and calculated bulk moduli (GPa) of 49 solids. 
  \label{tab:bmod}} \\
 \hline\hline
 Solid & Expt. & gKS-SCAN & FD-SCAN & PD-rSCAN \\
 \hline
 \endfirsthead
 \hline\hline
 Solid & Expt. & gKS-SCAN & FD-SCAN & PD-rSCAN \\
 \hline
 \endhead
 \hline
 \endfoot
 \hline\hline
 \endlastfoot
   Ag & 105.7 & 110.70 & 100.20 & 104.40 \\
   Al &  77.1 &  77.50 &  90.50 &  88.61 \\
 AlAs &  75.0 &  76.50 &  74.20 &  75.30 \\
  AlN & 206.0 & 212.10 & 206.20 & 216.00 \\
  AlP &  87.4 &  91.40 &  91.40 &  90.30 \\
   Au & 182.0 & 169.20 & 153.60 & 154.64 \\
   BN & 410.2 & 394.30 & 383.00 & 383.00 \\
   BP & 168.0 & 173.90 & 167.10 & 170.00 \\
   Ba &  10.6 &   8.30 &   9.90 &   7.21 \\
   Ca &  15.9 &  17.60 &  20.00 &  16.71 \\
   Cs &   2.3 &   1.90 &   2.40 &   1.15 \\
   Cu & 144.3 & 164.30 & 162.10 & 162.70 \\
 GaAs &  76.7 &  73.20 &  65.60 &  72.20 \\
  GaN & 213.7 & 194.10 & 183.30 & 250.00 \\
  GaP &  89.6 &  88.80 &  82.80 &  89.90 \\
   Ge &  79.4 &  71.20 &  66.70 &  68.56 \\
   Hf & 109.0 &      - &      - & 116.16 \\
 InAs &  58.6 &  57.80 &  50.50 &  59.60 \\
  InP &  72.0 &  68.90 &  65.50 &  66.10 \\
 InSb &  46.1 &  43.60 &  42.70 &  42.70 \\
   Ir & 362.2 & 407.20 & 357.00 & 374.20 \\
    K &   3.8 &   3.40 &   5.00 &   3.55 \\
   Li &  13.1 &  16.80 &  17.20 &  12.77 \\
 LiCl &  38.7 &  34.90 &  42.60 &  34.00 \\
  LiF &  76.3 &  77.90 &  83.20 &  81.30 \\
  LiH &  40.1 &  36.40 &  39.40 &  34.50 \\
   Mo & 276.2 & 275.30 & 270.30 & 272.99 \\
   Na &   7.9 &   8.00 &   8.90 &   6.90 \\
 NaCl &  27.6 &  28.70 &  32.00 &  27.00 \\
  NaF &  53.1 &  60.10 &  61.10 &  62.90 \\
   Nb & 173.2 & 177.10 & 180.40 & 171.90 \\
   Ni & 192.5 & 232.70 & 219.20 &      - \\
   Os & 418.0 &      - &      - & 425.01 \\
   Pd & 187.2 & 192.60 & 190.00 & 190.10 \\
   Pt & 285.5 & 291.80 & 249.60 & 271.10 \\
   Rb &   3.6 &   2.70 &   3.30 &   2.80 \\
   Re & 372.0 &      - &      - & 386.18 \\
   Rh & 277.1 & 293.50 & 254.40 & 281.24 \\
   Ru & 320.8 &      - &      - & 337.96 \\
   Sc &  39.4 &      - &      - &  56.68 \\
   Si & 101.3 &  99.70 &  94.40 &  96.90 \\
   Sr &  12.0 &  11.40 &  12.20 &  10.58 \\
   Ta & 202.7 & 208.20 & 201.00 & 197.35 \\
   Tc & 297.0 &      - &      - & 318.25 \\
   Ti & 105.1 &      - &      - & 117.67 \\
    V & 165.8 & 195.80 & 195.50 &      - \\
    W & 327.5 & 328.10 & 310.00 & 318.90 \\
    Y &  36.6 &      - &      - &  40.01 \\
   Zr &  83.3 &      - &      - &  95.52 \\
  \hline
\end{longtable}

\end{document}